# Experimental demonstration of cavity-free optical isolators and optical circulators

En-Ze Li,[1,2,*] Dong-Sheng Ding,[1,2,3,*,†] Yi-Chen Yu,[1,2] Ming-Xin Dong,[1,2] Lei Zeng,[1,2] Wei-Hang Zhang,[1,2] Ying-Hao Ye,[1,2] Huai-Zhi Wu,[4] Zhi-Han Zhu,[3] Wei Gao,[3] Guang-Can Guo,[1,2] and Bao-Sen Shi[1,2,‡]

[1]*Key Laboratory of Quantum Information, University of Science and Technology of China, Hefei, Anhui 230026, China*
[2]*Synergetic Innovation Center of Quantum Information and Quantum Physics, University of Science and Technology of China, Hefei, Anhui 230026, China*
[3]*Wang Da-Heng Collaborative Innovation Center for Science of Quantum Manipulation and Control, Heilongjiang Province and Harbin University of Science and Technology, Harbin 150080, China*
[4]*Fujian Key Laboratory of Quantum Information and Quantum Optics and Department of Physics, Fuzhou University, Fuzhou 350116, People's Republic of China*

Cavity-free optical nonreciprocity components, which have an inherent strong asymmetric interaction between the forward- and backward-propagation direction of the probe field, are key to produce such as optical isolators and circulators. According to the proposal presented by Xia *et al.*, [Phys. Rev. Lett. **121**, 203602 (2018)], we experimentally build a device that uses cross-Kerr nonlinearity to achieve a cavity-free optical isolator and circulator. Its nonreciprocal behavior arises from the thermal motion of *N*-type configuration atoms, which induces a strong chiral cross-Kerr nonlinear response for the weak probe beam. We obtain a two-port optical isolator for up to 20 dB of isolation ratio in a specially designed Sagnac interferometer. The distinct propagation directions of the weak probe field determine its cross-phase shift and transmission, by which we demonstrate the accessibility of a four-port optical circulator.

*Introduction.* Nonreciprocity optical devices that break the time-reversal symmetry are very difficult to achieve without magnetic fields, such as optical isolators and circulators [1]. To break reciprocity, the traditional method is to guide light through a medium with a powerful magneto-optical Faraday effect [2–4]. However, systems with this nature often have serious conflicts with miniaturization due to the surrounding environmental interference from their strong magnetic fields. The requirements of a nonmagnetic isolator have generated tremendous impetus, and then a series of works in various physical principles are reported to avoid using magneto-optical components. For example, the different types of optical nonreciprocal devices have been realized by using spatiotemporal modulation of nonlinear material [5–11], inducing the Berry phase [7,12–14], and using optomechanical systems [14–18].

To realize optical isolators with Kerr and Kerr-like nonlinearity is attracting many researchers [19–25]. However, optical isolators with Kerr and Kerr-like nonlinearity has a poor isolation effect under the weak signals due to the limitation of dynamic nonreciprocity [26,27]. Some nonlinear optical isolators with chiral gains are reported to overcome this difficulty in [28,29]. Besides, the nonlinear optical isolation schemes require a high-quality cavity to achieve enough interaction strength, such as optical resonators [21,28,30–33]. Although the fundamental aspects of optical nonreciprocity have been studied before, it is challenging to achieve an optical nonreciprocity with high isolation, low loss, and weak light intensity simultaneously. Therefore a passive nonlinear isolator without dynamic reciprocity would be of interest and is proposed by [32].

Here we demonstrate a cross-Kerr nonlinearity based on an *N*-type atomic configuration in a thermal vapor cell to achieve a cavity-free optical isolator and circulator. The cross-Kerr nonlinearity affects the transmission and susceptibility of the input weak probe field dramatically; this process depends on the coupling and switch fields' forward- and backward-propagating directions with respect to the probe field. Hence we achieve a cross-Kerr optical isolator with up to 20 dB of isolation ratio. In addition, we demonstrate a four-port optical circulator via a specially designed Sagnac interferometer. Compared to Ref. [28], they use the four-level nonreciprocal amplification mechanism, which creates an active process to break the dynamic nonreciprocity. On the contrary, we use the enhanced cross-Kerr nonlinearity, which is not limited between the classical regime and quantum regime. The reported cross-Kerr optical isolator and optical circulator could work under high isolation, low loss, and weak field, which holds potential applications for quantum communication [34],

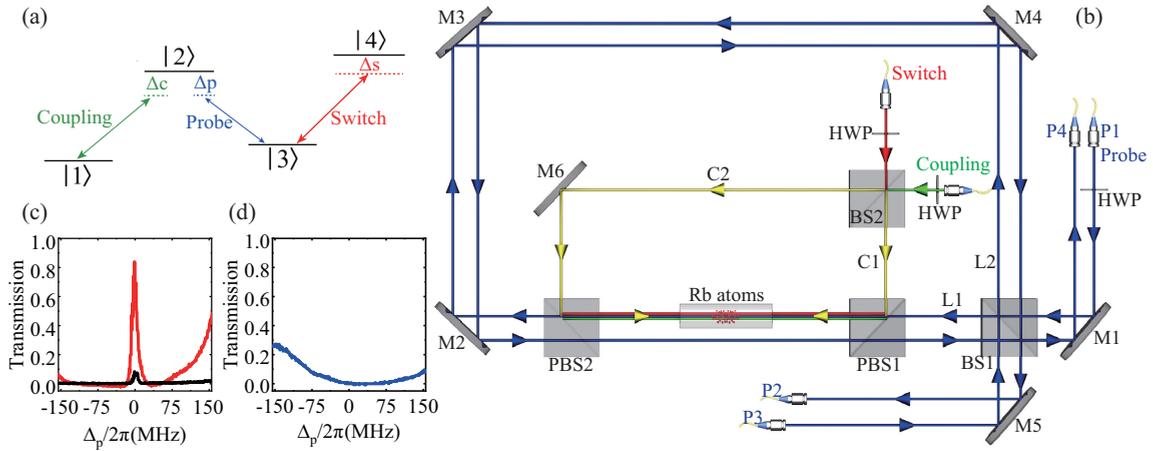

FIG. 1. (a) Energy diagram of optical isolation and circulation, corresponding to the *N*-type four states in $^{85}$Rb atoms (D line) of $5S_{1/2}(F = 2)$ ($|1\rangle$), $5P_{1/2}(F = 3)$ ($|2\rangle$), $5S_{1/2}(F = 3)$ ($|3\rangle$), and $5P_{3/2}(F = 3)$ ($|4\rangle$), respectively. (b) Schematic overview of the experimental setup, including a Sagnac interferometer system. The blue lines represent the probe beam path, red and green lines represent the switch and the coupling laser beams, respectively. M: optical mirrors; BS: beam splitters (50/50); PBS: polarizing beam splitters; HWP: half-wave plates; L1 and L2 are two different paths of the Sagnac interferometer, in which a Rb cell is located in L1; and C1 and C2 are the combined laser beams of switch field and coupling field. (c) The copropagation transmission vs the detuning of the probe field, the red curve and black curve recorded as the switch power $I_s = 21$ mW and $I_s = 0$ mW, respectively. (d) The counterpropagating (blue curve) transmission vs the detuning of the probe field, with the switch power $I_s = 21$ mW. In (c) and (d) the probe field is set as $I_p = 7.5$ µW and the coupling field set as $I_c = 12$ mW.

quantum simulation [35], and quantum information processing [36].

*System configurations.* We use an *N*-type atomic configuration in thermal rubidium ($^{85}$Rb) atoms to demonstrate the cross-Kerr nonlinearity [37–39]. As shown in Fig. 1(a), the coupling and switch lasers with vertical polarization couple the atomic transition $|5S_{1/2}, F = 2\rangle$ ($|1\rangle$) $\longleftrightarrow$ $|5P_{1/2}, F = 3\rangle$ ($|2\rangle$), and $|5S_{1/2}, F = 3\rangle$ ($|3\rangle$) $\longleftrightarrow$ $|5P_{3/2}, F = 3\rangle$($|4\rangle$), respectively. The probe laser with horizontal polarization couples the transition $|5P_{1/2}, F = 3\rangle$ ($|2\rangle$) $\longleftrightarrow$ $|5S_{1/2}, F = 3\rangle$ ($|3\rangle$). The Rabi frequencies and detunings of the probe, coupling, and switch fields are respectively denoted by $\Omega_p$ ($\Delta_p$), $\Omega_c$ ($\Delta_c$), and $\Omega_s$ ($\Delta_s$). The spontaneous decay rates of the state $|2\rangle$ ($|4\rangle$) to states $|1\rangle$ and $|3\rangle$ are $\Gamma_{21}$ ($\Gamma_{41}$) and $\Gamma_{23}$ ($\Gamma_{43}$), respectively. The dephasing rate between the two ground states $|1\rangle$ and $|3\rangle$ is $\Gamma_{31}$. In the rotating-wave approximation, the Hamiltonian under the interaction picture describing the field-atom interaction takes the form

$$H_I = \hbar(\Delta_c - \Delta_p)\sigma_{33} + \hbar\Delta_c\sigma_{22} + \hbar(\Delta_c - \Delta_p + \Delta_s)\sigma_{44}$$
$$- (\hbar/2)(\Omega_c\sigma_{12} + \Omega_s\sigma_{34} + \text{H.c.}),  \quad (1)$$

where $\sigma_{mn} = |m\rangle\langle n|$ ($m, n = 1, 2, 3, 4$) are the atomic transition operators. Due to atomic thermal motion, the frequencies of the propagating fields seen by the atoms are shifted, which is referred to as the directional Doppler effect. The detunings of the probe, coupling, and switch field are then modified as $\Delta_p \pm k_p v$, $\Delta_c \pm k_c v$, and $\Delta_s \pm k_s v$, where $k_p$, $k_c$, and $k_s$ are the corresponding wave vectors, and the signs "±" depend on the propagation direction. When the probe beams copropagate with the coupling and the switch field, the Doppler shifts take the same sign and their effects are generally canceled. However, when the probe beam counterpropagates with the coupling and the switch field, the Doppler shifts have opposite signs, and thus the Doppler broadening cannot be ignored. For both the co- and counterpropagation cases, we solve the steady-state solution of the master equation $d\rho/dt = -i[H, \rho]/\hbar - \mathcal{L}[\sqrt{\Gamma}\sigma_-]\rho$ by taking all the decay channels into consideration and obtain the total cross-Kerr nonlinear susceptibility averaged over the velocity distribution as

$$\chi_{23}^{\pm} = \int \rho_{23}^{\pm}(\Delta_p \pm k_p v, \Delta_c \pm k_c v, \Delta_s \pm k_s v, \Omega_p, \Omega_c, \Omega_s)$$
$$\times \frac{|\mu_{23}|^2}{\hbar\varepsilon_0}\frac{N(v)}{\Omega_p}dv, \quad (2)$$

where $N(v) = N_0 \exp(-v^2/u^2)/(u\sqrt{\pi})$, $u = (2k_BT/M)^{-1/2}$, $N_0$ is the atomic density, $\mu_{23}$ the transition dipole moment between states $|2\rangle$ and $|3\rangle$, $k_B$ is the Boltzmann constant, $T$ is the temperature of the gas in the cell, and M is the atomic mass. $\chi_{23}^{+}$ ($\chi_{23}^{-}$) represents the probe field copropagating (counterpropagation) with the coupling and the switch field. The transmission for the probe field is further given by the imaginary part of optical susceptibility Im[$\chi_{23}^{\pm}$], which strongly depends on the propagation direction of the probe field regarding the coupling and the switch field, leading to the chiral cross-Kerr nonlinearity. This effect allows us to implement optical isolators and circulators by handling the probe field with fixed directional coupling and switch fields.

As schematically illustrated in Fig. 1(b), the specially designed Sagnac interferometer consists of BS1 and three mirrors M2–M4, which has two different optical paths L1 and L2. This interferometer has four ports consisting of two input ports (P1 and P3) and two output ports (P2 and P4). We insert PBS1 and PBS2 into the interferometer and add a Rb cell in the optical path L1. To implement isolators and circulators, the most important part is a subsystem including five elements (BS2, M6, PBS1-2, and the Rb cell) which can control the transmission and the phase shift of the input probe beam. In this subsystem, the BS2 mixes the coupling and the switch beam into two collinear beams along with the optical directions of C1 and C2 simultaneously. The probe beam splits into two beams along L1 and L2 by the BS1 (beam splitter). Due to the chiral cross-Kerr nonlinearity, the probe beam accumulates different phase shifts while it propagates along path L1 and path L2, based on which we can implement a two-port optical isolator by closing the path L2, or implement an optical circulator with both the L1 and L2 paths being opened.

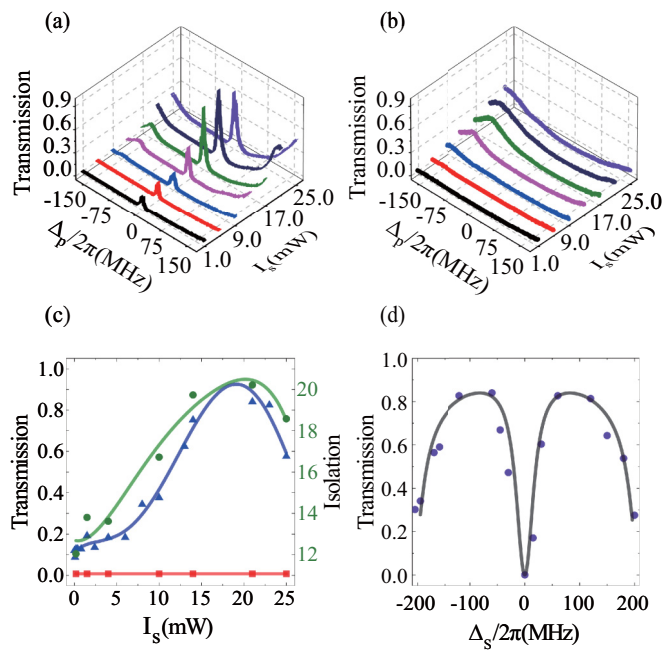

FIG. 2. Experimental observation of nonreciprocal transmission. (a), (b) The co- and counterpropagation transmission spectra of the probe beam by increasing switch field power $I_s$ with the probe field power $I_p = 7.5\,\mu$W, the coupling field power $I_c = 12$ mW, temperature $T = 38\,°$C. The beam diameter of the probe, coupling, and switch field are both 2 mm (c) The on-resonance transmission of the isolator for the copropagating (blue triangles) and counterpropagating (red squares), and the isolation ratio of the isolator (green dots). (d) The on-resonance transmission spectra of the probe field vs the detuning of the switch field (violet dots). In (c), (d) the dots are experimental results, and the solid lines are theoretical fittings.

*Isolator.* We input the weak probe field with the power $I_p = 7.5\,\mu$W to measure the co- or counterpropagation transmission. The copropagation (counterpropagation) transmission corresponds to opening C1 (C2) and blocking C2 (C1). The co- and counterpropagation transmissions are given in Figs. 1(c) and 1(d). By changing the switch power $I_s$ from 0 to 25 mW, we obtain a series of transmission spectra of the co- and counterpropagating probe fields in Figs. 2(a) and 2(b). The on-resonance transmission of the probe beam increases in the copropagation case but remains small for the counterpropagation case. When the combined coupling and switch beams C1 copropagate with the probe beam, the Doppler shift seen by the atoms has the same sign and their Doppler effect in the $N$-type configuration can be eliminated; thus we can observe a strong response of the probe field to the cross-Kerr nonlinearity. In contrast, the Doppler effect under the counterpropagation case strongly diminishes the cross-Kerr nonlinearity, so we observe that the probe transmission is almost vanishing. Although this $N$-type configuration and the theoretical $N$-type structure are slightly different, these two kinds of four-level systems equally investigate the cross-Kerr nonlinear effect. The nonreciprocity arising from the asymmetric cross-Kerr nonlinearity directly leads to the direction-dependent response of the probe beam, thus resulting in the nonreciprocal feature of the probe field.

We characterize the property of the isolator by measuring the resonant probe transmission with $\Delta_p = 0$ and by the logarithmic ratio of the copropagation transmission $T_{co}$ to the counterpropagation transmission $T_{cou}$, namely, $10 \log 10(T_{co}/T_{cou})$ (referred to as the isolation ratio), as shown in Fig. 2(c). There is a significant difference between the on-resonance transmission spectrum of the copropagation (blue dots) and counterpropagation (red dots). In the copropagation case, we observe that the increase of the switch field power $I_s$ can significantly enhance the cross-Kerr nonlinear response of the probe field. In the conventional three-level electromagnetically induced transparency (EIT) scheme, the nonlinear response susceptibility is relatively small. We add a switch field based on the $\Lambda$-type EIT scheme, and the absorption loss can be effectively reduced by increasing the intensity of the coupling field and the switch field. Such an $N$-type configuration system greatly enhances the cross-Kerr nonlinearity [40,41]. While the Doppler shift breaks the chiral cross-Kerr nonlinearity in the counterpropagation case, we obtain a series of weak on-resonance transmission signals. Moreover, the isolation ratio increases against $I_s$, as predicted by the theory in the article [32]. As the switch power increases, the isolation ratio increases to be 20 dB. Even though we scan the switch field power $I_s$ only from 0 to 25 mW because of the limitation of the laser power in our experiment, we nevertheless find that when the switch field power is tuned exceeding 21 mW, the transmission and the isolation ratio will be suppressed. This is because the atoms would be depleted when using a large power of the switch field to pump. Moreover, we further examine the copropagation transmission versus the detuning $\Delta_s$ of the switch field with the fixed switch field power $I_s = 21$ mW, as shown in Fig. 2(d). For a resonant pumping, i.e., $\Delta_s = 0$ MHz, the switch laser will suppress the cross-Kerr nonlinearity, while the off-resonant pumping of the switch field will induce the strong cross-Kerr nonlinear response. In addition, we observe that increasing the power of the switch field produces a magnifying effect on the copropagation transmission of the probe field. Remarkably, all the experimental data (dots) fit well with our theoretical prediction (dashed line), thus fully confirming our physical insights and theoretical analysis above.

*Circulator.* We unblock path L2 and enable the two noncoincident paths of the Sagnac interferometer to work simultaneously, which constructs a four-port optical circulator as proposed in [32]. We inject the probe beam into this circulator by P1 and control the propagation direction of the switch and the coupling beams with C1, C2. While the probe field copropagates with the other two laser beams, the strong cross-Kerr nonlinearity cannot only modify the transmission rate $T_{co}$ but also introduce a large phase shift $\phi_{co}^{L1}$ to the probe field in contrast to that of the counterpropagating case. The difference in phase shift for the two noncoincident paths $\Delta\Phi = \phi_{co}^{L1} - \phi^{L2}$ leads to the interference effect of the two counterpropagating probe fields at P2, which can be measured by the phase-dependent intensity fringes [42,43]. To achieve that, the beam splitter (BS) has two functions: first dividing the input probe field into two paths and then mixing them after passing through the Rb cell. If the reflection and transmission amplitudes of BS are $\sin\theta$ and $\cos\theta$ and the relative phase $\varphi$, then the BS yields the photon operation in the form of $\hat{a}_{out}^\dagger = \cos\theta \hat{a}_{in}^\dagger + ie^{-i\varphi}\sin\theta \hat{b}_{in}^\dagger$ and $\hat{b}_{out}^\dagger = ie^{i\varphi}\sin\theta \hat{a}_{in}^\dagger + \cos\theta \hat{b}_{in}^\dagger$ [42–44]. We choose $\theta = \pi/4$ (50:50 BS) and $\varphi = 0$ in the experiment.

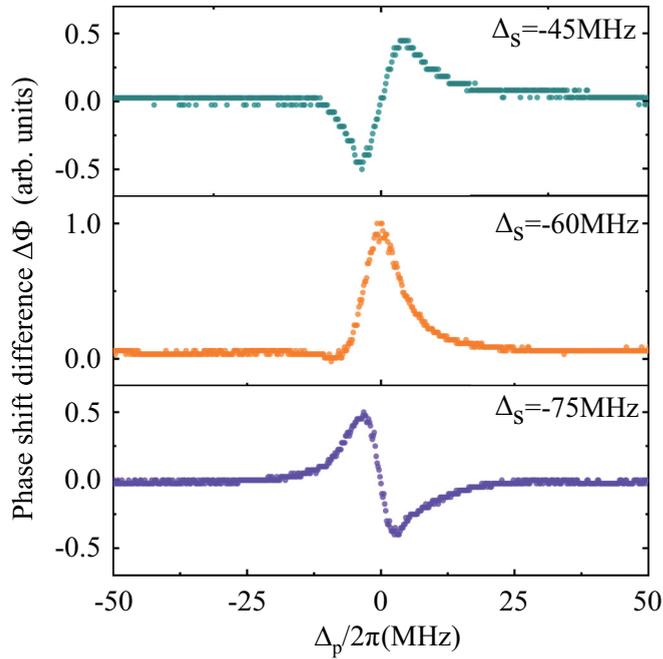

FIG. 3. Circulator performance. (a)–(c) The three different phase shifts vs the probe field detuning $\Delta_p$ with fixed switch field $I_s = 21$ mW, coupling field $I_c = 12$ mW, and probe field $I_p = 7.5$ μW.

The experimental results obtained previously tell us that the transmission of the co- and counterpropagation can be very different. Because of the cross-Kerr nonlinearity in the copropagation case, the transmission $T_{co}$ and the phase shift $\phi_{co}$ of the probe beam can be modified by the switch laser beam. We choose the best EIT transmission peak ($I_s = 21$ mW) to obtain the phase shift between path L1 and L2. The coupling field locked on the respective transition ($\Delta_c = 0$), and the switch field is on the off-resonance condition. As the detuning of the probe beam changes, we observe three different patterns of the phase-shift difference $\Delta\Phi$ by tuning the phase-shift offset between the optical paths L1 and L2. At a specific probe frequency, the phase-shift difference $\Delta\Phi = \phi_{co}^{L1} - \phi^{L2}$ between two opposite propagating probes can be measured by detecting the intensity of port 2. Under the condition of copropagation direction (open C1 and block C2), the data of the phase-shift difference $\Delta\Phi$ are shown in Figs. 3(a)–3(c).

The phase-shift properties in Figs. 3(a) and 3(c) are the property of the interferometry. To better understand the observation, we assume that the phase shift of the two arms in the Sagnac interferometer is caused by the nonlinear medium. The cross-Kerr nonlinearity induces a significant change in the refractive index of the medium, and the phase shift can be expressed as $\phi_{co}^{L1} = k_p \Delta n_1 L_1$ and $\phi^{L2} = k_p \Delta n_2 L_2$. According to our theoretical model, $L_1$ represents the length of the Rb cell, which is 10 cm, and we set the $\phi^{L2}$ as zero for simplicity, and the refractive index change $\Delta n_1 \propto \text{Re}[\chi_{23}]I_p$. We note that the phase shift between path L1 and L2 is caused by the refractive index change of the nonlinear medium, which is related to the real part of the susceptibility $\chi_{23}$. The imaginary part of the susceptibility $\chi_{23}$ induces the transmission change in the isolator part. To better understand this result, the relationship between the probe transmission and the refractive index governed by the Kramers-Kronig relations was found to be in excellent agreement with the observation data [43,45].

Figures 3(a) and 3(c) show the whole working window of the circulator. The phase-shift change supports our previous view that the scans show the detuning of the probe field near the working window of the circulator—the probe detuning induces the sensitive phase shift due to the enhanced cross-Kerr nonlinearity, at the top of the interference curve, which represents the output path $P1 \rightarrow P2$, and at the bottom of the curve, which represents the output path $P1 \rightarrow P4$. In this way, we succeed in realizing the fixed output path $P1 \rightarrow P2$ or path $P1 \rightarrow P4$ of the circulator. As is shown in Fig. 3(b), by changing the detuning of the switch field we observe an ideal phase-shift shape of the circulator, which only represents the path $P1 \rightarrow P2$. This clearly demonstrates that the circulator satisfies the output path $P1 \rightarrow P2$. Moreover, the amplitude of the phase shifts increases with the switch field power until it starts to saturate, but the phase-shift shape remains unchanged. The performance of the circulator can be quantified with the contrast $T = (T_{max}^i - T_{min}^i)/(T_{max}^i + T_{min}^i)$; here i represents the output port number. In our system, the circulator contrast can reach 0.902, while an ideal operation yields $T = 1$.

In addition, the contrast of the circulator is also limited by the quality of the Sagnac interferometer, and the measured contrast includes all loss and noise. According to the theoretical proposal [32], the phase difference between the two paths of the interferometer also satisfies P3 to P4. It is worth mentioning that our optical circulator can also work under a reversible case. If we turn on C2 and block C1, and we exchange the corresponding input port and output port completely, then we can get a circulator which is exactly the opposite of the previous port sequence. Then this apparatus can realize the circulation process of $P2 \rightarrow P3$ and $P4 \rightarrow P1$, implying that our specifically designed apparatus can be used as a four-port circulator, for example, a nonreciprocal photon circulation along with port $1 \rightarrow 2 \rightarrow 3 \rightarrow 4 \rightarrow 1$.

*Conclusion.* In summary, we have demonstrated an experiment to realize cavity-free optical isolators and circulators by a chiral Cross-Kerr nonlinearity of $N$-type Rb atoms embedded in the two-path Sagnac interferometer at room temperature. Simultaneously, our isolator can reach a high isolation ratio of 20 dB. Based on the experimental conclusions, we successfully prove that the specific designed experimental device can provide a new version for the optical circulator. Therefore our design allows us to make optical isolators and circulators for high isolation, low loss, and a weak probe field at room temperature.

The demonstrated circulator concept is useful for the processing and routing of the classical signals at weak light in classical optical circuits and quantum networks. The circulation operation principle is universal in a large variety of different quantum systems, as long as the medium in this system can reach enough optical depth to realize the EIT, and they can be miniaturized, for example, using the microcell [46–48], the atomic cladding waveguides [49], the hollow-core photonic crystal fibers [50–52], the slot waveguides [53], and some kinds of chip-based structures [48,54].

Arranging $N$ circulators as a linear array allows one to realize a $(2N+2)$-port optical circulator. A two- or three-dimensional network of optical circulators will be the potential candidate in multipath signal processing [30,55].

*Note added.* Recently, we found a relative work by Lin et al. [29] on nonreciprocal amplification with a four-level atomic system.

*Acknowledgments.* This work was supported by the National Key R&D Program of China (2017YFA0304800), the National Natural Science Foundation of China (Grants No. 61525504, No. 61722510, No. 61435011, No. 11604322, and No. 11934013), and the Innovation Fund from CAS, Anhui Initiative in Quantum Information Technologies (AHY020200), and the Youth Innovation Pro motion Association of Chinese Academy of Sciences under Grant No. 2018490. H.Z.W. acknowledges the support of the National Natural Science Foundation of China under Grant No. 11705030, the Natural Science Foundation of Fujian Province under Grant No. 2017J01401, and the Qishan fellowship of Fuzhou University.

[1] D. Jalas, A. Petrov, M. Eich, W. Freude, S. Fan, Z. Yu, R. Baets, M. Popović, A. Melloni, J. D. Joannopoulos *et al.*, What is – and what is not – an optical isolator, Nat. Photonics **7**, 579 (2013).

[2] H. Dötsch, N. Bahlmann, O. Zhuromskyy, M. Hammer, L. Wilkens, R. Gerhardt, P. Hertel, and A. F. Popkov, Applications of magneto-optical waveguides in integrated optics, JOSA B **22**, 240 (2005).

[3] M. Levy, Nanomagnetic route to bias-magnet-free, on-chip Faraday rotators, JOSA B **22**, 254 (2005).

[4] R. J. Potton, Reciprocity in optics, Rep. Prog. Phys. **67**, 717.

[5] Z. Yu and S. Fan, Optical isolation based on nonreciprocal phase shift induced by interband photonic transitions, Appl. Phys. Lett. **94**, 171116 (2009).

[6] H. Lira, Z. Yu, S. Fan, and M. Lipson, Electrically Driven Nonreciprocity Induced by Interband Photonic Transition on a Silicon Chip, Phys. Rev. Lett. **109**, 033901 (2012).

[7] L. D. Tzuang, K. Fang, P. Nussenzveig, S. Fan, and M. Lipson, Non-reciprocal phase shift induced by an effective magnetic flux for light, Nat. Photonics **8**, 701 (2014).

[8] D. L. Sounas and A. Alù, Non-reciprocal photonics based on time modulation, Nat. Photonics **11**, 774 (2017).

[9] Z. Yu and S. Fan, Complete optical isolation created by indirect interband photonic transitions, Nat. Photonics **3**, 91 (2009).

[10] M. S. Kang, A. Butsch, and P. S. J. Russell, Reconfigurable light-driven opto-acoustic isolators in photonic crystal fibre, Nat. Photonics **5**, 549 (2011).

[11] C. R. Doerr, L. Chen, and D. Vermeulen, Silicon photonics broadband modulation-based isolator, Opt. Express **22**, 4493 (2014).

[12] K. Fang, Z. Yu, and S. Fan, Realizing effective magnetic field for photons by controlling the phase of dynamic modulation, Nat. Photonics **6**, 782 (2012).

[13] K. Fang, Z. Yu, and S. Fan, Photonic Aharonov-Bohm Effect Based on Dynamic Modulation, Phys. Rev. Lett. **108**, 153901 (2012).

[14] K. Fang, J. Luo, A. Metelmann, M. H. Matheny, F. Marquardt, A. A. Clerk, and O. Painter, Generalized non-reciprocity in an optomechanical circuit via synthetic magnetism and reservoir engineering, Nat. Phys. **13**, 465 (2017).

[15] J. Chan, T. P. M. Alegre, A. H. Safavi-Naeini, J. T. Hill, A. Krause, S. Gröblacher, M. Aspelmeyer, and O. Painter, Laser cooling of a nanomechanical oscillator into its quantum ground state, Nature (London) **478**, 89 (2011).

[16] M. Hafezi and P. Rabl, Optomechanically induced non-reciprocity in microring resonators, Opt. Express **20**, 7672 (2012).

[17] Z. Shen, Y.-Li Zhang, Y. Chen, C.-L. Zou, Y.-F. Xiao, X.-B. Zou, F.-W. Sun, G.-C. Guo, and C.-H. Dong, Experimental realization of optomechanically induced non-reciprocity, Nat. Photonics **10**, 657 (2016).

[18] F. Ruesink, M.-A. Miri, A. Alu, and E. Verhagen, Nonreciprocity and magnetic-free isolation based on optomechanical interactions, Nat. Commun. **7**, 13662 (2016).

[19] L. Fan, J. Wang, L. T. Varghese, H. Shen, B. Niu, Y. Xuan, A. M. Weiner, and M. Qi, An all-silicon passive optical diode, Science **335**, 447 (2012).

[20] B. Peng, Ş. K. Özdemir, F. Lei, F. Monifi, M. Gianfreda, G. L. Long, S. Fan, F. Nori, C. M. Bender, and L. Yang, Parity–time-symmetric whispering-gallery microcavities, Nat. Phys. **10**, 394 (2014).

[21] N. Bender, S. Factor, J. D. Bodyfelt, H. Ramezani, D. N. Christodoulides, F. M. Ellis, and T. Kottos, Observation of Asymmetric Transport in Structures with Active Nonlinearities, Phys. Rev. Lett. **110**, 234101 (2013).

[22] M. Soljačić, C. Luo, J. D. Joannopoulos, and S. Fan, Nonlinear photonic crystal microdevices for optical integration, Opt. Lett. **28**, 637 (2003).

[23] L. Chang, X. Jiang, S. Hua, C. Yang, J. Wen, L. Jiang, G. Li, G. Wang, and M. Xiao, Parity–time symmetry and variable optical isolation in active–passive-coupled microresonators, Nat. Photonics **8**, 524 (2014).

[24] V. Grigoriev and F. Biancalana, Nonreciprocal switching thresholds in coupled nonlinear microcavities, Opt. Lett. **36**, 2131 (2011).

[25] I. V. Shadrivov, K. Y. Bliokh, Y. P. Bliokh, V. Freilikher, and Y. S. Kivshar, Bistability of Anderson Localized States in Nonlinear Random Media, Phys. Rev. Lett. **104**, 123902 (2010).

[26] Y. Shi, Z. Yu, and S. Fan, Limitations of nonlinear optical isolators due to dynamic reciprocity, Nat. Photonics **9**, 388 (2015).

[27] A. B. Khanikaev and A. Alu, Optical isolators: Nonlinear dynamic reciprocity, Nat. Photonics **9**, 359 (2015).

[28] S. Hua, J. Wen, X. Jiang, Q. Hua, L. Jiang, and M. Xiao, Demonstration of a chip-based optical isolator with parametric amplification, Nat. Commun. **7**, 13657 (2016).

[29] G. Lin, S. Zhang, Y. Hu, Y. Niu, J. Gong, and S. Gong, Nonreciprocal Amplification with Four-Level Hot Atoms, Phys. Rev. Lett. **123**, 033902 (2019).

[30] M. Scheucher, A. Hilico, E. Will, J. Volz, and A. Rauschenbeutel, Quantum optical circulator controlled by a single chirally coupled atom, Science **354**, 1577 (2016).

[31] S. Zhang, Y. Hu, G. Lin, Y. Niu, K. Xia, J. Gong, and S. Gong, Thermal-motion-induced non-reciprocal quantum optical system, Nat. Photonics **12**, 744 (2018).

[32] K. Xia, F. Nori, and M. Xiao, Cavity-Free Optical Isolators and Circulators Using a Chiral Cross-Kerr Nonlinearity, Phys. Rev. Lett. **121**, 203602 (2018).

[33] J. Kim, M. C. Kuzyk, K. Han, H. Wang, and G. Bahl, Non-reciprocal Brillouin scattering induced transparency, Nat. Phys. **11**, 275 (2015).

[34] N. Gisin and R. Thew, Quantum communication, Nat. Photonics **1**, 165 (2007).

[35] R. P. Feynman, Simulating physics with computers, Int. J. Theor. Phys. **21**, 467 (1982).

[36] A. N. Michael and Isaac L. Chuang, *Quantum Computation and Quantum Information* (Cambridge University Press, Cambridge, UK, 2011).

[37] H. Schmidt and A. R. Hawkins, Electromagnetically induced transparency in alkali atoms integrated on a semiconductor chip, Appl. Phys. Lett. **86**, 032106 (2005).

[38] M. Hofer, M. E. Fermann, F. Haberl, M. H. Ober, and A. J. Schmidt, Mode locking with cross-phase and self-phase modulation, Opt. Lett. **16**, 502 (1991).

[39] J. E. Sharping, M. Fiorentino, P. Kumar, and R. S. Windeler, All-optical switching based on cross-phase modulation in microstructure fiber, IEEE Photon. Technol. Lett. **14**, 77 (2002).

[40] H. Kang and Y. Zhu, Observation of Large Kerr Nonlinearity at Low Light Intensities, Phys. Rev. Lett. **91**, 093601 (2003).

[41] H. Schmidt and A. Imamoglu, Giant Kerr nonlinearities obtained by electromagnetically induced transparency, Opt. Lett. **21**, 1936 (1996).

[42] G. T. Purves, C. S. Adams, and I. G. Hughes, Sagnac interferometry in a slow-light medium, Phys. Rev. A **74**, 023805 (2006).

[43] G. Jundt, G. T. Purves, C. S. Adams, and I. G. Hughes, Non-linear Sagnac interferometry for pump-probe dispersion spectroscopy, Eur. Phys. J. D **27**, 273 (2003).

[44] P. Kok, W. J. Munro, K. Nemoto, T. C. Ralph, J. P. Dowling, and G. J. Milburn, Linear optical quantum computing with photonic qubits, Rev. Mod. Phys. **79**, 135 (2007).

[45] V. Lucarini, K. E. Peiponen, J. J. Saarinen, and E. M. Vartiainen, Kramers-Kronig relations in *optical materials research*, Springer Series in Optical Sciences Vol. 110 (Springer, New York, 2005).

[46] L. Weller, K. S. Kleinbach, M. A. Zentile, S. Knappe, I. G. Hughes, and C. S. Adams, An optical isolator using an atomic vapor in the hyperfine Paschen-Back regime, Opt. Lett. **37**, 3405 (2012).

[47] E. Talker, P. Arora, M. Dikopoltsev, and U. Levy, Optical isolator based on highly efficient optical pumping of Rb atoms in a miniaturized vapor cell, J. Phys. B: At., Mol. Opt. Phys. **53**, 045201 (2020).

[48] J. Kitching, Chip-scale atomic devices, Appl. Phys. Rev. **5**, 031302 (2018).

[49] L. Stern, B. Desiatov, I. Goykhman, and U. Levy, Nanoscale light–matter interactions in atomic cladding waveguides, Nat. Commun. **4**, 1548 (2013).

[50] P. S. Light, F. Benabid, F. Couny, M. Maric, and A. N. Luiten, Electromagnetically induced transparency in Rb-filled coated hollow-core photonic crystal fiber, Opt. Lett. **32**, 1323 (2007).

[51] A. D. Slepkov, A. R. Bhagwat, V. Venkataraman, P. Londero, and A. L. Gaeta, Spectroscopy of Rb atoms in hollow-core fibers, Phys. Rev. A **81**, 053825 (2010).

[52] K. Saha, V. Venkataraman, P. Londero, and A. L. Gaeta, Enhanced two-photon absorption in a hollow-core photonic-band-gap fiber, Phys. Rev. A **83**, 033833 (2011).

[53] R. Ritter, N. Gruhler, H. Dobbertin, H. Kübler, S. Scheel, W. Pernice, T. Pfau, and R. Löw, Coupling Thermal Atomic Vapor to Slot Waveguides, Phys. Rev. X **8**, 021032 (2018).

[54] R. Ritter, N. Gruhler, W. Pernice, H. Kübler, T. Pfau, and R. Löw, Atomic vapor spectroscopy in integrated photonic structures, Appl. Phys. Lett. **107**, 041101 (2015).

[55] R. Raussendorf and H. J. Briegel, A One-Way Quantum Computer, Phys. Rev. Lett. **86**, 5188 (2001).